\documentclass[conference]{IEEEtran}
\IEEEoverridecommandlockouts

\usepackage{soul}
\usepackage{xparse}
\usepackage{array}
\usepackage{amsmath}
\usepackage{amssymb}
\usepackage{amsthm}
\newtheorem{assumption}{Assumption}
\newtheorem{definition}{Definition}
\newtheorem{remark}{Remark}

\usepackage{textcomp}
\usepackage{stfloats}
\usepackage{lettrine}
\usepackage{multirow}
\usepackage{lipsum}
\usepackage{algorithm}
\usepackage{algpseudocode}
\usepackage{booktabs}
\usepackage{enumitem}
\usepackage[switch]{lineno}
\usepackage{cite}
\usepackage{graphicx}
\usepackage{xcolor}
\usepackage{makecell}
\usepackage{url}
\usepackage[hidelinks]{hyperref}
\usepackage{cleveref}

\definecolor{sinaBlue}{RGB}{0,76,153}
\definecolor{aliGreen}{RGB}{0,130,70}
\definecolor{rouzOrange}{RGB}{220,110,0}
\definecolor{marcusPurple}{RGB}{120,60,160}
\definecolor{revRed}{RGB}{200,40,40}
\definecolor{compactYellow}{RGB}{255,242,140}

\newcommand{\Revision}[1]{\textcolor{revRed}{#1}}

\usepackage{caption}
\begin{document}

\title{A Stackelberg–Bayesian Capacity-Market Game of Carbon Regulation and Second-Life Battery Investment under AI Data-Center Load Growth}

\author{
\IEEEauthorblockN{
Rouzbeh Haghighi$^{\dagger}$,
Ali Hassan$^{\dagger}$,
Sina Mohammadi$^{\dagger}$,
Marcus Chen I Wada,
and Wencong Su$^{*}$\\
}
\IEEEauthorblockA{
\textit{Dept. of Electrical and Computer Engineering}\\
\textit{University of Michigan--Dearborn}\\
Dearborn, MI, USA}

\thanks{$^{\dagger}$R. Haghighi, A. Hassan, and S. Mohammadi are co-first authors with equal contributions.}
\thanks{$^{*}$Corresponding author: Wencong Su (e-mail: wencong@umich.edu).}
}

\IEEEaftertitletext{\vspace{-3\baselineskip}}

\maketitle

\begin{abstract}
Artificial intelligence (AI) data centers are driving rapid electricity load growth across all U.S. ISO/RTO regions, raising both system costs and carbon exposure. This study develops a three-level Stackelberg--Bayesian game in which a regulator (leader) sets carbon penalties and subsidies, a single ISO capacity market clears against an energy balance modeled as a classical generation-expansion problem, and technology-specific investors (followers) decide capacity and operation under incomplete information, yielding a Bayesian Nash equilibrium. The AI impact is captured parsimoniously as an additional load-growth factor on a greenfield-incremental expansion, isolating how much new capacity the growth pulls in and which technology fills it. Within this framework, we consider second-life battery (SLB) storage competing against new/first-life storage for capacity-market revenue. We quantify how a carbon tax, a renewable subsidy, and an SLB subsidy reshape the equilibrium investment mix, carbon emissions, and profit. Different scenarios are compared at the end based on cost-effectiveness and reduced carbon emissions.
\end{abstract}


\section{Introduction}
Data centers have become a dominant driver of long-term electricity demand. This shift is propelled by GPU-accelerated artificial-intelligence (AI) and generative-AI workloads \cite{vercellino2026measurement, mohammadi2026grid}. After nearly fifteen years of flat consumption, U.S. data-center load rose from roughly 76\,TWh in 2018 to 176\,TWh in 2023, about 4.4\% of national use \cite{shehabi2024usdatacenter}. Near-term projections expect this load to double or triple by 2028 \cite{doe2024datacenterdemand}. Such growth forces large ISO regions to plan and build substantial new capacity over the next decade. At the same time, federal and state carbon policy pushes those additions toward clean resources \cite{eia2026aeo}.

Meeting this demand cleanly is not simply a matter of choosing a socially preferred technology mix and building it. In the stylized ISO capacity-market setting considered here, private investors develop capacity in response to expected energy and capacity revenues. Policymakers influence those incentives only indirectly, through instruments such as carbon taxes and technology-specific subsidies. The central question is therefore whether a given policy makes the socially preferred mix privately profitable at market equilibrium. Stackelberg formulations suit such hierarchical decisions: prior work provisions data-center resources against cost-minimizing followers \cite{yang2016stackelberg}, casts a storage operator as leader to a data-center follower under carbon policy \cite{zhang2025stackelberg}, and routes inference workloads across geo-distributed centers to a carbon- and cost-minimizing Nash equilibrium \cite{hogade2024game}. These formulations optimize short-run workload placement; none address the upstream generation and storage investment that sustained AI-driven demand growth requires.

Energy storage is a critical enabler of this growth, from chip-level buffering to grid-interactive uninterruptible power supply systems \cite{mohammadi2026grid}. New-BESS procurement in the U.S.\ is constrained by geopolitical supply chains and by demand that outpaces production. Second-life batteries (SLBs) offer a complementary path. Packs retired from electric vehicles (EVs) can no longer meet transportation performance requirements, but they still retain 70--80\% of their original capacity \cite{hassan2023second, haghighi2025deep}. They are cheaper, lower in embodied carbon, and faster to deploy for stationary grid and data-center support\Revision{.} SLBs nonetheless carry penalties: shorter life, faster fade, and limited feedstock. Whether investors actually build SLBs therefore depends on the carbon tax, the available subsidies, and how SLBs compete against new batteries for capacity-market revenue. A projected surplus of 100--200\,GWh of retired EV batteries by 2030, together with technoeconomic evidence that their grid-storage value exceeds their recycling value, positions retired packs as a low-cost, rapidly available resource \cite{Zhuang2025SLB}. This makes them an attractive match for the capacity-adequacy needs of AI data centers, which new purpose-built storage may struggle to meet on the same timeline.

Carbon policy and capacity-market design jointly shape these long-term build decisions. A bilevel generation-expansion-planning (GEP) framework has been used to study renewable-portfolio policy, with subsidy-driven renewable investment at the upper level and wholesale market clearing at the lower level \cite{nguyen2020generation}. That model reproduces effects such as merit-order price suppression and credit-price compression. We adopt a related leader--follower structure, but we shift the followers from renewable generators under credit subsidies to storage investors choosing between first- and second-life batteries under a carbon tax and AI-data-center load growth. Prior work has treated the relevant pieces separately: game-theoretic GEP under carbon policy \cite{haghighi2021generation}, storage participation in capacity markets, and AI-data-center load studies have not been combined. To our knowledge, no prior study couples an SLB entrant, AI-DC-driven load growth, a regulator-led hierarchy, and incomplete information in a single capacity-market game. We close this gap with a three-level Stackelberg--Bayesian capacity-market game: a regulator leader $\rightarrow$ a capacity market $\rightarrow$ investor followers under incomplete information. The model pairs a physically interpretable AI-DC load model with a throughput-based SLB model. It quantifies whether carbon penalties and subsidies shift the equilibrium toward second-life storage. The main contributions are: I) a regulator-led three-level game that prices capacity for an SLB-vs-new-battery investment choice; II) an AI-DC load captured parsimoniously as a demand-growth factor on a greenfield-incremental expansion; III) an incomplete-information (Bayesian Nash) lower level that quantifies the value of information for investors; and IV) battery end-of-life (EOL) tracking, so that first-life batteries (FB) and SLBs compete over their true service lives.

\vspace{-2mm}
\section{System Model and Market Framework}
\vspace{-2mm}
The market is a three-level game. A regulator (leader) commits to a public policy vector $\boldsymbol{\theta}=\{\tau_t,\sigma^{\mathrm{re}}_t, \sigma^{\mathrm{sl}}_t\}$ (carbon tax, renewable subsidy, SLB subsidy); a single-region capacity market clears subject to an energy balance in the form of a classical generation-expansion model; and investors choose capacity and dispatch to maximize profit. Planning years are indexed by $t\in\{1,\dots,T\}$, representative operating periods by $h\in\mathcal{H}$ with weights $w_h$ (period durations, $\sum_h w_h{=}8760$); a year is discounted by $\delta_t=(1+r)^{-t}$, where $r>0$ is the annual discount rate. The pre-existing fleet is excluded under a \emph{greenfield benchmark} (\cref{sec:dc}), only the AI-driven incremental demand is served, isolating how much new capacity that growth pulls in and which technology fills it. The setting is a single aggregate region rather than a named market, so the framework transfers to any large U.S. ISO/RTO footprint. Policy originates on the regulatory (RTO) side and enters every player's problem as public data; the ISO is a purely technical, non-strategic entity that clears the market and returns $(\pi_t,\lambda_{t,h})$; followers are technology-specific investors forming a Bayesian Nash equilibrium (BNE) under incomplete information. We reserve ``RTO'' for the policy side and ``ISO'' for the clearing side throughout.

\vspace{-4mm}
\subsection{ISO market and players}
\vspace{-2mm}
The investor set $\mathcal{P}=\mathcal{R}\cup\mathcal{G}\cup\mathcal{S}$ comprises renewable-eligible technologies $\mathcal{R}$ (solar PV, wind, biomass; eligible for $\sigma^{\mathrm{re}}$, other generators $\mathcal{G}$ (zero-emission nuclear, gas, CCHP; fossil subset $\mathcal{G}^{f}$), and storage $\mathcal{S}=\{\mathrm{FB},\mathrm{SLB}\}$ (FB: new BESS; SLB: second-life BESS). Each investor owns one technology and chooses integer builds $x^{m}_{t}\in\mathbb{Z}\ge0$, giving \emph{live} capacity $\bar K^{m}_{t}=\mathrm{CU}^{m}\sum_{\tau=\max(1,t-L_m+1)}^{t}x^{m}_{\tau}$ (only units within service life $L_m$, \cref{sec:slb}) and dispatch $P^{m}_{t,h}$. The market settles an energy price $\lambda_{t,h}$ (dual of the energy balance) and a capacity price $\pi_{t}$ (a cost-of-new-entry (CONE) curve, \cref{sec:clearing}). The game is run primarily in the capacity market, where the central question is whether SLBs earn enough capacity and energy revenue to be built.

\vspace{-2mm}
\subsection{AI data-center load model}
\label{sec:dc}
For the multi-year generation-expansion game the data center is not a strategic player; its aggregate footprint enters demand through an additional AI-DC load-growth factor. The cumulative demand-growth multiplier is \mbox{$\phi_t =\prod_{\tau=1}^{t}(1+g_{\tau}+\alpha^{\mathrm{AIDC}}_{\tau})$}, where \mbox{$g_t$} is the baseline growth rate and \mbox{$\alpha^{\mathrm{AIDC}}_t$} the incremental AI-DC factor. Peak and energy demands and their greenfield-incremental components follow as

\vspace{-8mm}
{\small
\begin{align}
\hat D_t&{=}\hat D_0\phi_t,\quad D^E_{t,h}{=}D^E_{0,h}\phi_t,\nonumber\\
\Delta\hat D_t&{=}\hat D_0(\phi_t{-}1),\quad \Delta D^E_{t,h}{=}D^E_{0,h}(\phi_t{-}1)
\label{eq:aidc_incremental_demand}
\end{align}}
\vspace{-6mm}

\noindent Under the greenfield benchmark the existing fleet serves the base-year load, while new capacity is built only to supply the incremental demand $\Delta \hat D_t$ and $\Delta D^E_{t,h}$.

\vspace{-2mm}
\subsection{Second-life battery model}
\label{sec:slb}
\vspace{-1mm}
The usable energy capacity of a second-life battery degrades with cumulative throughput. With initial capacity $E^{0}_{\mathrm{SLB}}$, state of health $\mathrm{SoH}_t$, and degradation coefficient $\alpha^{\mathrm{fade}}$.
\vspace{-4mm}

{\small
\begin{align}
E^{\mathrm{SLB}}_t &= E^{0}_{\mathrm{SLB}}\,\mathrm{SoH}_t \\
\mathrm{SoH}_t &= \mathrm{SoH}_0-\alpha^{\mathrm{fade}}\,Q_t \\
Q_t &=\!\sum_{\tau\le t}\sum_{h} w_h\big(P^{\mathrm{ch}}_{\tau,h}+P^{\mathrm{dis}}_{\tau,h}\big)
\label{eq:slb_soh}
\end{align}}
\vspace{-2mm}

\noindent where $Q_t$ is the cumulative energy throughput. The per-year cost combines annualized repurposing investment, throughput degradation with charging energy, and O\&M; the discounted lifetime cost is $C^{s}{=}\sum_t\delta_t C^{s}_t$:
{\small
\begin{subequations}
\label{eq:slb_cost}
\renewcommand{\theequation}{\theparentequation\alph{equation}}
\begin{align}
C^{s}_t
&= C^{\mathrm{Inv}}_{s,t}
+ C^{\mathrm{Gen}}_{s,t}
+ C^{\mathrm{O\&M}}_{s,t}
\label{eq:slb_cost_total}\\
C^{\mathrm{Inv}}_{s,t}
&= \mathrm{CRF}_s\left(C^{\mathrm{acq}}_t+\mathrm{RPC}\right)D_s\,\bar P^{s}_{t}
\label{eq:slb_cost_inv}\\
C^{\mathrm{Gen}}_{s,t}
&= \textstyle\sum_{h} w_h\lambda_{t,h}P^{\mathrm{ch}}_{s,t,h}
+ c^{\mathrm{deg}}\!\textstyle\sum_{h} w_h\big(P^{\mathrm{dis}}_{s,t,h}{+}P^{\mathrm{ch}}_{s,t,h}\big)
\label{eq:slb_cost_gen}\\
C^{\mathrm{O\&M}}_{s,t}
&= O^{s}\,\bar P^{s}_{t}
\label{eq:slb_cost_om}\\
\delta_t 
&= (1+r)^{-t}
\label{eq:slb_discount}
\end{align}
\end{subequations}}
\vspace{-6mm}

\noindent where $\mathrm{RPC}$ is the repurposing cost (testing, transport, sorting, repackaging, BMS upgrade, installation), $D_s$ the storage duration, and $C^{\mathrm{Inv}}_{s,t}$ the annualized capex (over $L_s$ via $\mathrm{CRF}_s$) charged on live power $\bar P^{s}_{t}$. A battery reaches EOL at $L_s=\min(L^{\mathrm{cal}}_s,L^{\mathrm{cyc}}_s)$, the smaller of calendar and cycle life. Because SLBs enter already aged ($\mathrm{SoH}_0\in[0.7,0.9]$), their life is short (5--8\,yr) versus 10--15\,yr for FB, so an SLB built early may retire and require replacement within the horizon. Capacity is therefore tracked by build-year \emph{vintage}: a unit built in year $\tau$ contributes only over $[\tau,\tau{+}L_s{-}1]$, its capex is annualized over $L_s$ via the capital-recovery factor $\mathrm{CRF}_s$, and a salvage value $\rho_s$ is recovered at retirement. Replacement is endogenous: if adequacy \cref{eq:adequacy} still binds after a vintage retires, the expansion rebuilds, charging each technology over its true service life for a fair FB-vs-SLB comparison. Here $L_s{=}L^{\mathrm{cal}}_s$ with an aggregate $\mathrm{SoH}$.

{\setlength{\abovecaptionskip}{-0.1mm}
\setlength{\belowcaptionskip}{-7mm}
\footnotesize
\begin{figure*}[!t]
  \centering
  \includegraphics[width=1\textwidth]{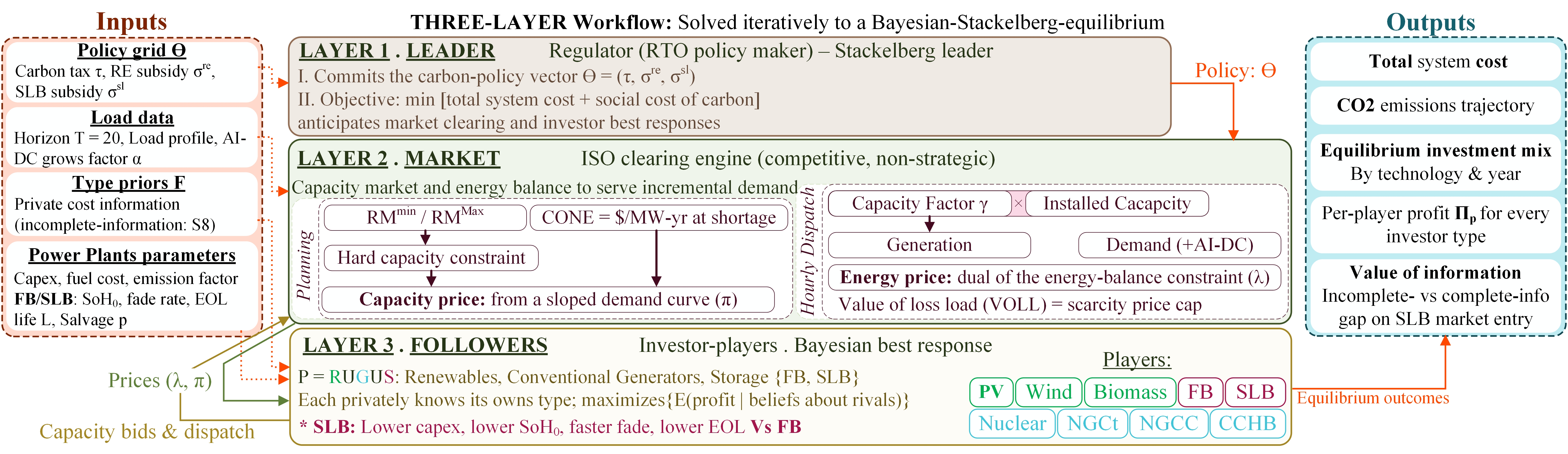}
  \caption{Three-Layer Stackelberg-Bayesian Game for Second-Life Battery Investment.}
  \label{fig:workflow}
\end{figure*}}

\vspace{-3mm}
\section{Stackelberg--Bayesian Game Formulation}
\vspace{-1mm}
\subsection{Proposed method}
\vspace{-1mm}
Each investor maximizes the present value of profit $\Pi^{p}{=}\sum_t\delta_t\Pi^{p}_t$ by choosing yearly capacity builds and dispatch; annualized investment uses the capital-recovery factor $\mathrm{CRF}^{m}$, and $\mathrm{cc}^{m}\in[0,1]$ is the resource-adequacy credit of technology $m$. The prices $(\lambda,\pi)$ settled in \cref{sec:clearing} are the two revenue channels every player faces. \Cref{fig:workflow} summarizes how the three layers interact to reach an equilibrium under a given policy $\boldsymbol{\theta}$. \\
\emph{\textbf{Layer~1} (leader):} the regulator commits $\boldsymbol{\theta}$ before investors act.\\
\emph{\textbf{Layer~2} (market):} serves the greenfield incremental demand $\Delta\hat D_t$ and clears $(\lambda,\pi)$ per \cref{sec:clearing} \cref{eq:balance,eq:cone}. Carbon enters only through the tax in each emitting payoff \cref{eq:gen,eq:re}, detailed in \cref{sec:clearing}. \\
\emph{\textbf{Layer~3} (followers):} the investors $\mathcal{R}\cup\mathcal{G}\cup\mathcal{S}$, each privately knowing its cost type $\vartheta_p$ and playing a Bayesian best response; SLB competes against FB at the same prices despite a lower cost and harsher aging profile (\cref{sec:slb}). Best responses are solved year-by-year under a continuous-build relaxation, warm-started from the planner benchmark and rounded to modules \mbox{$\mathrm{CU}^{m}$}. \\
\emph{\textbf{Solution loop:}} For a fixed $\boldsymbol{\theta}$, an inner diagonalization loop computes the equilibrium investment mix (\cref{sec:eqm}); the outer loop applies this across the predefined policy scenarios of \cref{tab:Results}, recording the equilibrium mix, system cost, CO$_2$ emissions, and the value of information (\cref{rem:voi}).

\vspace{-3mm}
\subsection{Player payoffs}
\vspace{-1mm}
A generator $g\in\mathcal{G}$ (emission factor $\varepsilon_g$; nuclear $\varepsilon_g{=}0$) earns
\vspace{-7mm}

{\small
\begin{align}
\Pi^{g}_t =\;&\textstyle\sum_{h} w_h\,\lambda_{t,h}P^{g}_{t,h}
 + \pi_{t}\,\mathrm{cc}^{g}\bar K^{g}_{t}
 \nonumber\\
 &-\textstyle\sum_{h} w_h\big(c^{g}+\tau_t\varepsilon_g\big)P^{g}_{t,h}
 -\big(\mathrm{CRF}^{g}I^{g}\mathrm{CU}^{g}x^{g}_{t}+O^{g}\bar K^{g}_{t}\big)
\label{eq:gen}
\end{align}}
\vspace{-5mm}

\noindent with $0\le P^{g}_{t,h}\le\bar K^{g}_{t}$ (CCHP belongs to $\mathcal{G}$ with a high-efficiency, low effective $\varepsilon$). A renewable-eligible developer $r\in\mathcal{R}$ receives the subsidy on its capex,

\vspace{-5mm}
{\small
\begin{align}
\Pi^{r}_t =\;&\textstyle\sum_{h} w_h\,\lambda_{t,h}P^{r}_{t,h}
 + \pi_{t}\,\mathrm{cc}^{r}\bar K^{r}_{t}
 -\textstyle\sum_{h} w_h\big(c^{r}+\tau_t\varepsilon_r\big)P^{r}_{t,h}
 \nonumber\\
 &-\big[(1-\sigma^{\mathrm{re}}_t)\mathrm{CRF}^{r}I^{r}\,\mathrm{CU}^{r}x^{r}_{t}
 +O^{r}\bar K^{r}_{t}\big]
\label{eq:re}
\end{align}}
\vspace{-4mm}

\noindent with $0\le P^{r}_{t,h}\le\gamma^{r}_{t,h}\bar K^{r}_{t}$ and capacity factor $\gamma^{r}$. A storage operator $s\in\{\mathrm{FB},\mathrm{SLB}\}$ earns discharge and capacity revenue against the investment, throughput, and O\&M costs of \cref{sec:slb} (charging cost $\textstyle\sum_{h} w_h\lambda_{t,h}P^{\mathrm{ch}}_{s,t,h}$ sits in $C^{\mathrm{Gen}}_{s}$). FB receives no build subsidy; SLB uses second-life parameters, is eligible for the build subsidy $\sigma^{\mathrm{sl}}_t$, and ages via \cref{eq:slb_soh}. The capacity term \mbox{$\pi_{t}\mathrm{cc}^{s}\mathrm{SoH}^{s}_{t}\bar P^{s}_{t}$} is the dominant value channel under flat AI-DC growth, and whether it (plus energy revenue and the SLB subsidy $\sigma^{\mathrm{sl}}_t$) covers the SLB cost stack is the question studied here:

\vspace{-4mm}
{\small
\begin{align}
\Pi^{s}_t =\;&\textstyle\sum_{h} w_h\,\lambda_{t,h}P^{\mathrm{dis}}_{s,t,h}
 + \pi_{t}\,\mathrm{cc}^{s}\,\mathrm{SoH}^{s}_{t}\bar P^{s}_{t}
 \nonumber\\
 &-\big((1{-}\sigma^{\mathrm{sl}}_t)C^{\mathrm{Inv}}_{s,t}+C^{\mathrm{Gen}}_{s,t}+C^{\mathrm{O\&M}}_{s,t}\big)
\label{eq:stor}
\end{align}}

\vspace{-4mm}
\subsection{Capacity market and energy balance}
\label{sec:clearing}
For a fixed strategy profile the market clears each period and the constraint duals define the prices, following the energy-balance and reserve-margin accounting of a classical generation-expansion model. New units serve the \emph{incremental} demand \cref{eq:aidc_incremental_demand}; the energy balance (dual $\lambda_{t,h}$) is

\vspace{-4mm}
{\small
\begin{align}
\textstyle\sum_{g}P^{g}_{t,h}+\sum_{r}P^{r}_{t,h}
 +\sum_{s}\!\big(P^{\mathrm{dis}}_{s,t,h}-P^{\mathrm{ch}}_{s,t,h}\big)
 +\,\mathrm{ns}_{t,h}=\Delta D^{E}_{t,h}
\label{eq:balance}
\end{align}}
\vspace{-5mm}

\noindent where $\mathrm{ns}$ is non-served energy priced at the value of lost load (VOLL). VOLL is a scarcity price cap, not a planning target: adequacy is enforced as a hard band \eqref{eq:adequacy} in the planner benchmark and incentivized through \mbox{$\pi_t$} \eqref{eq:cone} in the game. Let the live firm capacity be $\mathrm{CC}_t=\sum_{m}\mathrm{cc}^{m}\bar K^{m}_{t}+\sum_{s}\mathrm{cc}^{s}\mathrm{SoH}^{s}_{t}\bar P^{s}_{t}$. The \emph{planner benchmark} imposes the resource-adequacy band as a hard constraint,

\vspace{-4mm}
{\small
\begin{align}
(1{+}\mathrm{RM}^{\min}_t)\,\Delta\hat D_{t}\le \mathrm{CC}_t \le(1{+}\mathrm{RM}^{\max}_t)\,\Delta\hat D_{t}
\label{eq:adequacy}
\end{align}}
\vspace{-4mm}

\noindent with $\Delta\hat D_{t}$ the incremental coincident peak \cref{eq:aidc_incremental_demand}. In the \emph{market and the game}, capacity is instead priced by a CONE demand curve evaluated at $\mathrm{CC}_t$,

\vspace{-4mm}
{\small
\begin{align}
\pi_{t}=\mathrm{CONE}\cdot\mathrm{clip}\!\left(\frac{(1{+}\mathrm{RM}^{\max}_t)\Delta\hat D_{t}-\mathrm{CC}_t}{(\mathrm{RM}^{\max}_t-\mathrm{RM}^{\min}_t)\Delta\hat D_{t}},\,0,\,1\right)
\label{eq:cone}
\end{align}}
\vspace{-4mm}

\noindent so $\pi_t=\mathrm{CONE}$ when capacity is short of the adequacy target, slopes down through the band, and reaches $0$ in surplus. No system-wide carbon cap is imposed; each emitting unit (any $\varepsilon_g{>}0$, including biomass in $\mathcal{R}$) pays \mbox{$\tau_t\varepsilon_g$} per unit energy in \eqref{eq:gen} and \eqref{eq:re}, so the tax enters the offer and the clearing price \mbox{$\lambda_{t,h}$}. Unlike the planner, the game imposes no \mbox{$\mathrm{RM}^{\min}_t$} on any single player; adequacy is incentivized solely through \mbox{$\pi_t$}, a soft target whose equilibrium reserve margin may differ from the planner's.

\vspace{-2mm}
\subsection{Incomplete information}
\vspace{-2mm}
Each investor $p$ privately knows only its own cost \emph{type} $\vartheta_p\in\Theta_p$ and holds beliefs $b_p$ over rivals' types $\vartheta_{-p}$.
\vspace{-2mm}

\begin{assumption}[Incomplete information, common prior]
\label{as:info}
Types are drawn independently from a commonly known prior $F=\prod_p F_p$, so $b_p(\vartheta_{-p})= \prod_{q\ne p}F_q(\vartheta_q)$, and the policy $\boldsymbol{\theta}$ is public.
\end{assumption}
\vspace{-2mm}

Each player chooses a type-contingent strategy $s_p(\vartheta_p)$ maximizing its \emph{expected} discounted profit

\vspace{-4mm}
\begin{equation}
\max_{s_p(\cdot)}\;\; \mathbb{E}_{b_p}\!\Big[\textstyle\sum_t \delta_t\,
\Pi^{p}_t(\vartheta_p,\vartheta_{-p})\Big],
\label{eq:exp}
\end{equation}
\vspace{-6mm}

\noindent the expectation over rivals' types because clearing prices, and hence $p$'s revenue, depend on the realized profile; each player thus faces endogenous price uncertainty.
\vspace{-2mm}

{\setlength{\abovecaptionskip}{-0.1mm}
\setlength{\belowcaptionskip}{-4mm}
\footnotesize
\begin{table*}[!b]
\centering
\caption{Generation and storage technology parameters.}
\label{tab:params_generation_storage}
\renewcommand{\arraystretch}{1}
\setlength{\tabcolsep}{2pt}
\resizebox{\textwidth}{!}{%
\begin{tabular}{@{}c@{\hspace{1.5mm}}c@{}}
\begin{tabular}{@{}l c c c c c c c c c c@{}}
\toprule
\multicolumn{11}{@{}l}{\textit{Generation technologies}} \\
\midrule
Tech & Player & $\mathrm{CU}^{m}$ & $I$ & Fuel & O\&M & CF & $\varepsilon$ & $cc$ & $L^{\mathrm{cal}}$ & $\rho$ \\
     & set & (MW) & (\$/kW) & (\$/MWh) & (\$/kW-yr) & & (tCO$_2$/MWh) & & (yr) & (\$/MW) \\
\midrule
Wind    & R & 200     & 1{,}500 & 0  & 40  & 0.35 & 0.00 & 0.32 & 25 & 0   \\
Solar   & R & 200     & 1{,}200 & 0  & 20  & 0.22 & 0.00 & 0.12 & 30 & 0   \\
Biomass & R & 100     & 3{,}500 & 15 & 50  & 0.70 & 0.20 & 0.75 & 25 & 0   \\
Nuclear & G & 1{,}000 & 6{,}000 & 8  & 100 & 0.92 & 0.00 & 0.95 & 40 & 50k \\
NGCC    & G & 400     & 1{,}000 & 25 & 15  & 0.55 & 0.40 & 0.85 & 30 & 30k \\
NGCT    & G & 200     & 600     & 35 & 10  & 0.30 & 0.50 & 0.90 & 25 & 10k \\
CCHP    & G & 200     & 1{,}200 & 20 & 20  & 0.75 & 0.25 & 0.80 & 25 & 20k \\
\bottomrule
\end{tabular}
&
\begin{tabular}{@{}l c c c@{}}
\toprule
\multicolumn{4}{@{}l}{\textit{Storage technologies}} \\
\midrule
Parameter & Unit & FB & SLB \\
\midrule
Energy capex $C^{\mathrm{acq}}$ & \$/kWh & 280 & 80 \\
O\&M & \$/kW-yr & 15 & 12 \\
Degradation cost $c^{\mathrm{deg}}$ & \$/MWh & 2.0 & 4.0 \\
Round-trip efficiency $\eta_{\mathrm{rt}}$ & -- & 0.9025 & 0.7744 \\
Initial SoH $\mathrm{SoH}_0$ / floor $\underline{\mathrm{SoH}}$ & -- & 1.00 / 0.80 & 0.80 / 0.60 \\
Fade coefficient $\alpha^{\mathrm{fade}}$ & -- & 0.02 & 0.05 \\
Capacity credit $cc$ & -- & 0.85 & 0.70 \\
Salvage value $\rho_s$ & \$/kWh & 5 & 0 \\
\bottomrule
\end{tabular}
\end{tabular}}
\end{table*}}

\begin{definition}[Bayesian Nash equilibrium]
\label{def:bne}
For a fixed $\boldsymbol{\theta}$, a profile $\{s_p^\star(\cdot)\}$ and induced prices form a BNE if (i) for every $p$ and type $\vartheta_p$, $s_p^\star(\vartheta_p)$ maximizes \cref{eq:exp} given $b_p$ and $s_{-p}^\star(\cdot)$ subject to $p$'s payoff \cref{eq:gen,eq:re,eq:stor}; (ii) the energy balance \cref{eq:balance} holds for every realized type profile; and (iii) the energy price equals its dual and the capacity price follows the CONE curve \cref{eq:cone}.
\end{definition}
\vspace{-3mm}

\begin{remark}[Value of information]
\label{rem:voi}
The complete-information Nash point is the degenerate-belief case $b_p=\delta_{\vartheta_{-p}}$; the gap between the BNE and that point isolates how private information shifts SLB deployment timing and scale, not the terminal build-or-not outcome.
\end{remark}
\vspace{-2mm}

\vspace{-2mm}
\subsection{Equilibrium concept: from Nash to Bayesian Nash}
\label{sec:eqm}
\vspace{-2mm}
For a fixed policy $\boldsymbol{\theta}$, the equilibrium is found by an inner best-response diagonalization loop \cite{haghighi2021generation}. Each round: (i)~every player best-responds with rivals' builds fixed; (ii)~the ISO re-clears to update the prices $(\lambda,\pi)$; and (iii)~each player moves only partway to its best response, updating its build at rate $\omega\in(0,1]$ for stability. The loop repeats until no player gains more than tolerance $\epsilon$. Unlike \cite{haghighi2021generation}, $\lambda$ and $\pi$ move with every build, so the market is re-cleared each round rather than settled at a fixed price.

Under complete information (CI: S1--S7 and S9 in \cref{tab:Results}) the loop yields a pure-strategy Nash equilibrium consistent with degenerate beliefs; under incomplete information (II: S8 in \cref{tab:Results}) the inner step instead maximizes sample-average profit over $N_\xi$ type draws \cref{eq:exp}, with the no-deviation test applied per type, yielding the BNE \cref{def:bne}. The two nest: $N_\xi{=}1$ at the true types recovers the Nash loop exactly (a validation check), and the gap between the two equilibria is the value of information for SLB deployment timing and scale \cref{rem:voi}.

{\setlength{\abovecaptionskip}{-0.1mm}
\setlength{\belowcaptionskip}{-2mm}
\footnotesize
\begin{table*}[!b]
\centering
\caption{Scenario design and equilibrium outcomes by scenario (2026--2045).}
\label{tab:Results}
\renewcommand{\arraystretch}{1.15}
\setlength{\tabcolsep}{2pt}
\resizebox{\textwidth}{!}{%
\begin{tabular}{@{}c@{\hspace{1.5mm}}c@{}}
\begin{tabular}{@{}c l c l c c@{}}
\toprule
\multicolumn{4}{@{}l}{\textit{Scenario design}} \\
\midrule
\makecell{Sc.\\ No.} & \makecell{Scenario\\Name} & \makecell{Load\\$g/\alpha^{\mathrm{AIDC}}$} & \makecell{Regulator policy\\$\boldsymbol{\theta}$} & \makecell{Storage\\$\mathrm{FB/SLB}$} & \makecell{Game\\Info.} \\
\midrule
S1 & Baseline          & Base   & None              & None      & CI \\
S2 & Load growth       & AI-DC  & None              & None      & CI \\
S3 & Carbon tax        & AI-DC  & Carbon Tax (CT)   & None      & CI \\
S4 & First-life storage & AI-DC & CT                & FB        & CI \\
S5 & SLB investment    & AI-DC  & CT                & SLB       & CI \\
S6 & RE subsidy        & AI-DC  & CT + RE sub.      & SLB       & CI \\
S7 & SLB subsidy       & AI-DC  & CT + SLB sub.     & SLB       & CI \\
S8 & Incomplete info   & AI-DC  & CT + SLB sub.     & SLB       & II \\
S9 & FB vs SLB         & AI-DC  & CT                & FB + SLB  & CI \\
\bottomrule
\end{tabular}
&
\begin{tabular}{@{}c c c c c c | c c c c c | c@{}}
\toprule
\multicolumn{11}{@{}l}{\textit{Equilibrium outcomes}} \\
\midrule
$\Delta\hat D_{T}$ & Cost      & Carbon emission & $\bar{\pi}$ & Peak SLB & Peak FB & \multicolumn{5}{c}{2045 fleet (GW)} & $\Delta\varepsilon^{\mathrm{emb}}$ \\
%
(GW)               & (NPV-\$B) & (NPV-MtCO$_2$)  & (k\$/MW-yr) &  (GW) &  (GW) & Wind & Solar & Fossil & Nuclear & Total & (kt) \\
\midrule
39.6 & 29.74 & 300.1 & 77.3  & 0.0  & 0.0 & 1.60  & 0.0  & 61.40 & 0 & 63.0  & -- \\
62.2 & 54.30 & 543.7 & 75.4  & 0.0  & 0.0 & 2.80  & 0.0  & 96.20 & 0 & 99.0  & -- \\
62.2 & 72.58 & 420.8 & 73.3  & 0.0  & 0.0 & 10.80 & 2.80 & 93.80 & 0 & 107.4 & -- \\
62.2 & 70.92 & 376.8 & 57.1  & 0.0  & 2.2 & 8.80  & 4.20 & 92.00 & 2 & 107.0 & 0 \\
62.2 & 69.88 & 363.3 & 57.6  & 3.1  & 0.0 & 13.00 & 3.60 & 90.60 & 2 & 109.2 & 176.7 \\
62.2 & 49.61 & 91.9 & 55.2 & 2.6 & 0.0 & 67.40 & 41.40 & 64.20 & 0 & 173.0 & 148.2 \\
62.2 & 71.02 & 376.4 & 58.9  & 3.5  & 0.0 & 9.00  & 3.60 & 92.00 & 2 & 106.6 & 199.5 \\
62.2 & 75.76 & 458.1 & 158.4 & 16.2 & 0.0 & 19.40 & 28.40 & 83.40 & 0 & 131.2 & 974.7 \\
62.2 & 71.09 & 383.9 & 61.9  & 2.4  & 2.2 & 9.40  & 6.60 & 92.00 & 1 & 109.0 & 136.8 \\
\bottomrule
\end{tabular}
\end{tabular}}
\end{table*}}

\vspace{-2mm}
\section{Case Study and Results}
\vspace{-3mm}
\subsection{Test system}
The study uses a single aggregate ISO region over a 20-year horizon (2026--2045, $T{=}20$). Base demand grows at $g{=}1.5\%$/yr. In the AI-DC scenarios, an additional $\alpha^{\mathrm{AIDC}}{=}1.5\%$/yr is applied over 2026--2035 and set to zero thereafter, so AI-DC load plateaus from 2035. Post-2035 expansion is therefore largely replacement-driven, which exercises the battery EOL dynamics. In the baseline scenario (S1), $\alpha^{\mathrm{AIDC}}{=}0$ throughout. A discount rate of $r{=}5\%$ is used for all net-present-value (NPV) calculations. The base-year coincident peak is $\hat D_0{=}121{,}000$\,MW and base-year annual energy is $D^E_0{=}650$\,TWh/yr. Market parameters are $\mathrm{RM}^{\min}{=}15\%$, $\mathrm{RM}^{\max}{=}40\%$, $\mathrm{VOLL}{=}\$5{,}000$/MWh, and $\mathrm{CONE}{=}\$250{,}000$/MW-yr. Cost and performance parameters follow \cite{haghighi2021generation} and are summarized in \cref{tab:params_generation_storage}. Three quantities distinguish technology contribution: $\gamma_h$ ($\equiv\gamma^{r}_{t,h}$ \cref{eq:re}) is the achievable output fraction per period $h$; the capacity factor caps annual energy; and $cc$ is the nameplate fraction counted toward the reserve margin. Conventional units are fully dispatchable ($\gamma_h{=}1$) though they average only ${\sim}55\%$ output; renewables take $\gamma_h\in[0,1]$ from time-of-day availability, detailed in \cite{Haghighi2026_Git_Repo}. Biomass is RE-subsidy eligible but, carrying $\varepsilon{=}0.20$, still pays the carbon tax like any emitting units.

FB and SLB differ in energy capex, $\mathrm{SoH}_0$, degradation rate, and EOL service life, following the fair-market-value framework of \cite{Bach2025} (\cref{tab:params_generation_storage}). Both use $\mathrm{CU}^{m}{=}100$\,MW modules and $D{=}4$\,h duration, with efficiencies $\eta^{c}{=}\eta^{d}{=}0.95$ (FB) and $0.88$ (SLB); SLB adds a repurposing cost $\mathrm{RPC}{=}\$45$/kWh on top of energy capex \cref{eq:slb_cost_inv}. Calendar lives are $7$\,yr (SLB) and $12$\,yr (FB). The $\overline{FS}{=}60$\,GWh/yr feedstock cap (MISO's retired-EV-pack share of U.S.\ load) is non-binding in baseline runs---economics, not availability, limits deployment. Each MWh of SLB capacity avoids $\Delta\varepsilon^{\mathrm{emb}}{=}14.3$\,tCO\mbox{$_2$}e of embodied manufacturing carbon versus new packs \cite{dai2019life,peters2017environmental,kamath2020evaluating}, worth under 1\% of its acquisition cost at the modeled carbon tax.

\vspace{-2mm}
\subsection{Scenario design}
\vspace{-1mm}
\Cref{tab:Results} defines nine scenarios (S1--S9) spanning load growth (\emph{base}: $g$ only; \emph{AI-DC}: $g+\alpha^{\mathrm{AIDC}}$), storage eligibility, and information structure, under a carbon tax $\tau{=}50$\,\$/tCO\mbox{$_2$}, a renewable subsidy $\sigma^{\mathrm{re}}{=}0.25$, and an SLB subsidy $\sigma^{\mathrm{sl}}{=}0.25$ (fractions of annualized capex; no emissions cap, so carbon is priced only through $\tau$). Key contrasts: S4 vs.\ S5 (new vs.\ second-life storage), S5 vs.\ S7 (direct SLB subsidy), S6 vs.\ S7 (indirect RE vs.\ direct SLB subsidy), S7 vs.\ S8 (value of information), and S9 ($\sigma^{\mathrm{sl}}{=}0$: FB and SLB compete on economics alone).

\vspace{-2mm}
\subsection{Results and Discussion}
\label{sec:results}
\vspace{-2mm}
\Cref{tab:Results} reports the equilibrium outcomes. NPV cost and NPV CO$_2$ are discounted by $\delta_t$, with NPV CO$_2$ counting operational emissions only; Peak SLB/FB is the maximum live storage over $T$; the 2045 fleet lists live capacity by resource; and $\Delta\varepsilon^{\mathrm{emb}}$\ is the cumulative embodied CO$_2$ avoided by SLB against an all-new-BESS (FB). Load growth is first-order: the AI-DC overlay (S1$\to$S2) lifts the 2045 incremental peak from 39.6 to 62.2\,GW (${+}57\%$) and, before any policy, raises NPV CO$_2$ from 300 to 544\,Mt (${+}81\%$) and NPV cost by 83\%. A carbon tax alone (S3) trims emissions 23\% versus S2 but is the second-costliest case (\$72.6\,B). Indirect renewable support then dominates every direct storage policy: the RE subsidy (S6) delivers the lowest NPV cost of any AI-DC scenario (\$49.6\,B) and the lowest NPV CO$_2$ in the study (91.9\,Mt---below even the no-growth baseline), building 67.4\,GW wind and 41.4\,GW solar by 2045, an 83\% emissions cut under S2. The direct SLB subsidy (S7) does not help: despite a 25\% investment subsidy it leaves no live storage at 2045 and is marginally costlier and dirtier than unsubsidized SLB (S5, \$69.9\,B / 363\,Mt).

Second-life storage is a lifecycle, not a terminal-stock, outcome. Live storage is 0\,GW in 2045 across every storage-eligible scenario (S4--S9), yet SLB and FB enter transiently, peaking at 2.2--3.5\,GW mid-horizon under complete information. Mean cleared prices of $\bar\pi{\approx}55$--62\,k\$/MW-yr fall short of the revenue needed to rebuild batteries after calendar retirement, so the tax and repurposing-cost advantage shift \emph{when} SLB enters, not whether it survives the horizon. Incomplete information (S8) amplifies this transient---SLB peaks at 16.2\,GW and avoids 975\,kt of embodied CO$_2$---but the fleet still empties by 2045, while $\bar\pi$ jumps to 158\,k\$/MW-yr under repeated CONE-cap episodes. Head-to-head competition (S9, $\sigma^{\mathrm{sl}}{=}0$) confirms the mechanism: FB and SLB both deploy mid-horizon (peaks 2.4 and 2.2\,GW) but neither persists, avoiding 137\,kt of embodied CO$_2$ on economics alone. Carbon pricing by itself is thus insufficient to sustain SLB and, where subsidies do act, they clear the capacity market at lower prices ($\bar\pi{=}55$\,k\$/MW-yr in S6).

{\setlength{\abovecaptionskip}{-0.1mm}
\setlength{\belowcaptionskip}{-7mm}
\footnotesize
\begin{figure}[!t]
  \centering
  \includegraphics[width=0.8\columnwidth]{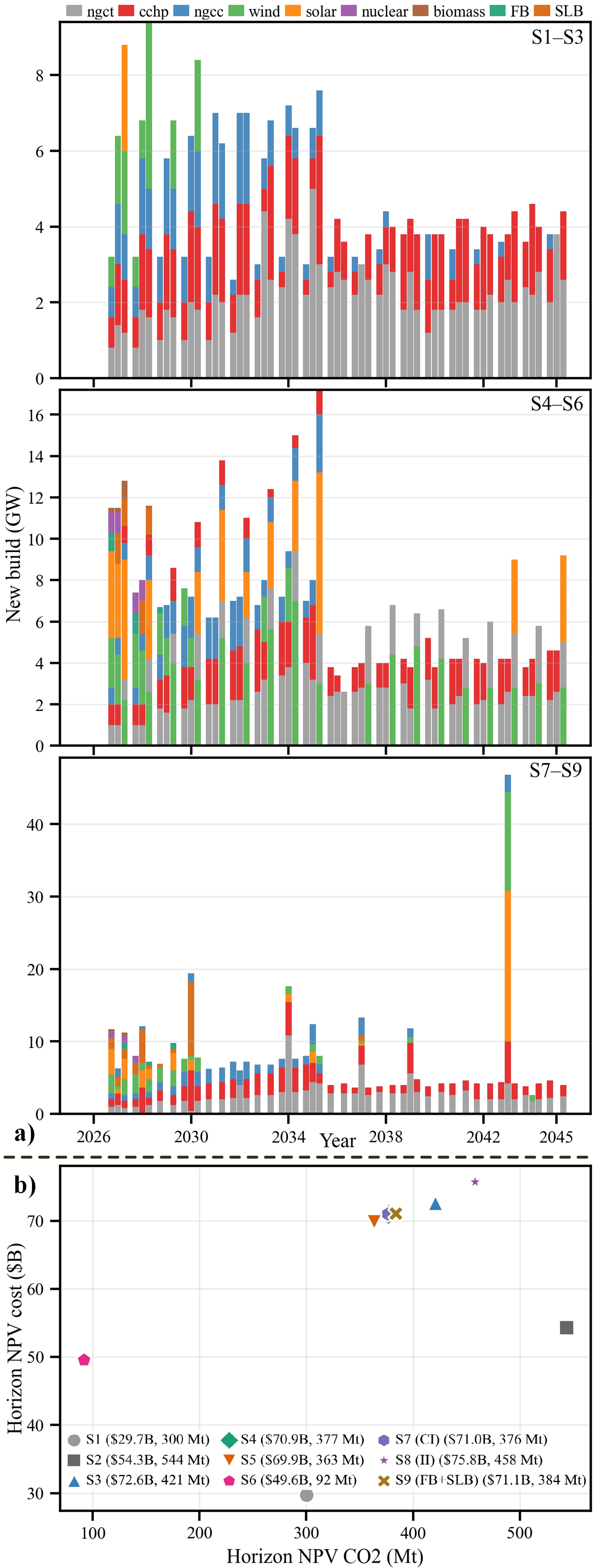}
  \caption{Equilibrium outcomes under different policy scenarios. \textbf{(a)}~Annual equilibrium new builds (GW) by technology; panels group S1--S3, S4--S6, and S7--S9 (three scenarios side-by-side per year). Subsidised wind and solar enter in the first half of the horizon; later builds are dominated by fossil replacement (NGCT, CCHP, NGCC). \textbf{(b)}~Horizon NPV cost versus operational NPV CO$_2$.}
  \label{fig:results}
\end{figure}}

Equilibrium uniqueness was verified on S6 and S7 at 2035 and 2045: the symmetric pseudo-gradient Jacobian $\tfrac{1}{2}(G{+}G^{\top})$ (central differences, $\delta{=}50$\,MW) is negative definite, satisfying Rosen's diagonal strict concavity~\cite{rosen1965existence} on the sloped CONE segment~\cref{eq:cone}, and $50$ randomized restarts of the damped iteration~\cref{sec:eqm} converge to an identical build profile ($0$\,MW deviation). Outcomes thus rest on a single Nash point; where $\pi_t$ saturates the technology split may not be unique, but total capacity credit, $\pi_t$, $\lambda_{t,h}$, and emissions are (proof in~\cite{Haghighi2026_Git_Repo}).

\Cref{fig:results} highlights the same intertemporal logic from two perspectives. The panels group scenarios \(\mathrm{S1}\)--\(\mathrm{S3}\), \(\mathrm{S4}\)--\(\mathrm{S6}\), and \(\mathrm{S7}\)--\(\mathrm{S9}\); within each year, the bars are ordered from left to right by scenario. The equilibrium new-build paths front-load wind, solar, and, where eligible, storage in the first decade because discounted operating and capacity-market revenues, together with any investment subsidy collected at commissioning, must be earned over the remaining horizon. A late installation captures fewer subsidy-linked and market rents before 2045 and therefore cannot compete with early entry in the Nash equilibrium.

The figure also reports the horizon NPV cost and operational NPV CO$_2$ emissions for \(\mathrm{S1}\)--\(\mathrm{S9}\). AI-DC load growth, from \(\mathrm{S1}\) to \(\mathrm{S2}\), shifts the outcome sharply upward and to the right, increasing cost from \(\$29.7\,\mathrm{B}\) to \(\$54.3\,\mathrm{B}\) and emissions from \(300\,\mathrm{Mt}\) to \(544\,\mathrm{Mt}\). With carbon tax and storage policies active, most scenarios, \(\mathrm{S3}\)--\(\mathrm{S5}\) and \(\mathrm{S7}\)--\(\mathrm{S9}\), fall within a tight high-cost band of \(\$70\)--\(\$76\,\mathrm{B}\). In contrast, the renewable-energy subsidy case, \(\mathrm{S6}\), reaches \(\$49.6\,\mathrm{B}\) and \(92\,\mathrm{Mt}\), yielding both lower cost and lower emissions than any other AI-DC case. Scenario \(\mathrm{S8}\), with incomplete information, is the most expensive point in the cluster, while \(\mathrm{S5}\) and \(\mathrm{S7}\) marginally reduce CO$_2$ emissions relative to \(\mathrm{S3}\) but cannot match the performance of \(\mathrm{S6}\).

\vspace{-2mm}
\section{Conclusion}
\vspace{-2mm}
This paper formulated a three-level Stackelberg--Bayesian capacity-market game to study whether SLB are built when AI data-center load grows, carbon is priced, and RE or SLB subsidies are available. On a stylized 20-year ISO generation expansion, AI-DC load is the dominant driver of incremental adequacy needs and emissions. A renewable investment subsidy (S6) sharply decarbonizes the incremental build at lower cost than a comparable SLB subsidy (S7) or carbon tax alone (S3). Under default policy, SLB enters transiently but does not persist to 2045 in any complete-information scenario; short calendar life and soft capacity-market incentives make replacement, not initial entry, the binding constraint. Incomplete information (S8) shifts deployment timing and nearly quintuples peak SLB scale relative to the matched complete-information case (S7, 16.2 vs.\ 3.5\,GW), yet the zero terminal-stock outcome is unchanged. Future work should couple these market incentives with explicit EV-battery feedstock constraints and regulator welfare objectives over the cost--emissions frontier.

\vspace{-3mm}
\footnotesize
\bibliographystyle{IEEEtran}
\bibliography{Ref}

@misc{Haghighi2026_Git_Repo,
  author       = {Rouzbeh Haghighi and Ali Hassan and Sina Mohammadi and Marcus Chen I Wada and Wencong Su},
  year         = {2026},
  url = {https://github.com/RouzbehHaghighi/SLB_AIDC_Game},
  note         = {GitHub repository}
}

@techreport{shehabi2024usdatacenter,
  author      = {Shehabi, Arman and Smith, Sarah J. and Hubbard, Alex and Newkirk, Anna and Lei, Nuoa and Siddik, Md Abu Bakar and Holecek, Brian and Koomey, Jonathan G. and Masanet, Eric R. and Sartor, Dale A.},
  title       = {{2024 United States Data Center Energy Usage Report}},
  institution = {Lawrence Berkeley National Laboratory (LBNL)},
  year        = {2024},
  url         = {https://escholarship.org/uc/item/32d6m0d1}
}

@misc{doe2024datacenterdemand,
  author       = {{U.S. Department of Energy}},
  title        = {{DOE Releases New Report Evaluating Increase in Electricity Demand from Data Centers}},
  year         = {2024},
  url          = {https://www.energy.gov/articles/doe-releases-new-report-evaluating-increase-electricity-demand-data-centers},
}

@article{vercellino2026measurement,
  title={Measurement of Generative AI Workload Power Profiles for Whole-Facility Data Center Infrastructure Planning},
  author={Vercellino, Roberto and Willard, Jared and Campos, Gustavo and Pereira, Weslley da Silva and Hull, Olivia and Selensky, Matthew and Mueller, Juliane},
  journal={arXiv preprint arXiv:2604.07345},
  year={2026}
}

@techreport{Bach2025,
  author      = {A. Bach and S. Onori and S. J. Reichelstein and J. Zhuang},
  title       = {Fair Market Value of Used Capacity Assets: Forecasts for Repurposed Electric Vehicle Batteries},
  institution = {ZEW Discussion Paper No. 25-065},
  year        = {2025},
  url         = {https://hdl.handle.net/10419/333927}
}

@misc{eia2026aeo,
  author = {{U.S. Energy Information Administration}},
  title = {{Annual Energy Outlook 2026}},
  year = {2026},
  month = apr,
}

@article{haghighi2021generation,
  title={Generation expansion planning using game theory approach to reduce carbon emission: A case study of Iran},
  author={Haghighi, Rouzbeh and Yektamoghadam, Hossein and Dehghani, Majid and Nikoofard, Amirhossein},
  journal={Computers \& Industrial Engineering},
  volume={162},
  pages={107713},
  year={2021},
  publisher={Elsevier}
}

@article{mohammadi2026grid,
  title={Grid Integration of AI Data Centers: A Critical Review of Energy Storage Solutions},
  author={Mohammadi, Sina and Wang, Wayne and Wada, Marcus Chen I and Haghighi, Rouzbeh and Hassan, Ali and Liu, Hualong and Bhatnagar, Archit and Chen, Ang and Su, Wencong},
  journal={arXiv preprint arXiv:2603.00415},
  year={2026}
}

@article{hassan2023second,
  title={Second-life batteries: A review on power grid applications, degradation mechanisms, and power electronics interface architectures},
  author={Hassan, Ali and Khan, Shahid Aziz and Li, Rongheng and Su, Wencong and Zhou, Xuan and Wang, Mengqi and Wang, Bin},
  journal={Batteries},
  volume={9},
  number={12},
  pages={571},
  year={2023},
  publisher={MDPI}
}

@article{kamath2020evaluating,
  title={Evaluating the cost and carbon footprint of second-life electric vehicle batteries in residential and utility-level applications},
  author={Kamath, Dipti and Shukla, Siddharth and Arsenault, Renata and Kim, Hyung Chul and Anctil, Annick},
  journal={Waste Management},
  volume={113},
  pages={497--507},
  year={2020},
  publisher={Elsevier},
  doi={10.1016/j.wasman.2020.05.034}
}

@article{dai2019life,
  title={Life Cycle Analysis of Lithium-Ion Batteries for Automotive Applications},
  author={Dai, Qiang and Kelly, Jarod and Gaines, Linda and Wang, Michael},
  journal={Batteries},
  volume={5},
  number={2},
  pages={48},
  year={2019},
  publisher={MDPI},
  doi={10.3390/batteries5020048}
}

@article{peters2017environmental,
  title={The environmental impact of Li-Ion batteries and the role of key parameters -- A review},
  author={Peters, Jens F. and Baumann, Manuel and Zimmermann, Benedikt and Braun, Joachim and Weil, Marcel},
  journal={Renewable and Sustainable Energy Reviews},
  volume={67},
  pages={491--506},
  year={2017},
  publisher={Elsevier},
  doi={10.1016/j.rser.2016.08.039}
}

@inproceedings{haghighi2025deep,
  title={Deep reinforcement learning-based optimization of second-life battery utilization in electric vehicles charging stations},
  author={Haghighi, Rouzbeh and Hassan, Ali and Bui, Van-Hai and Hussain, Akhtar and Su, Wencong},
  booktitle={2025 IEEE Power \& Energy Society General Meeting (PESGM)},
  pages={1--5},
  year={2025},
  organization={IEEE}
}

@article{Zhuang2025SLB,
  title   = {Technoeconomic decision support for second-life batteries},
  author  = {Zhuang, Jihan and Bach, Amadeus and van Vlijmen, Bruis H. C.
             and Reichelstein, Stefan J. and Chueh, William and Onori, Simona
             and Benson, Sally M.},
  journal = {Applied Energy},
  volume  = {390},
  pages   = {125800},
  year    = {2025},
  doi     = {10.1016/j.apenergy.2025.125800}
}

@article{zhang2025stackelberg,
  title   = {A Stackelberg-game based bi-level scheduling model of data center
             combined with shared energy storage considering price linkage and
             demand response},
  author  = {Zhang, Shuo and Wei, Ming and Li, Yingzi and Chen, Yuanli},
  journal = {Energy},
  volume  = {336},
  pages   = {138509},
  year    = {2025},
  doi     = {10.1016/j.energy.2025.138509}
}

@article{yang2016stackelberg,
  title={Stackelberg game approach for energy-aware resource allocation in data centers},
  author={Yang, Bo and Li, Zhiyong and Chen, Shaomiao and Wang, Tao and Li, Keqin},
  journal={IEEE Transactions on Parallel and Distributed systems},
  volume={27},
  number={12},
  pages={3646--3658},
  year={2016},
  publisher={IEEE}
}

@article{hogade2024game,
  title={Game-theoretic deep reinforcement learning to minimize carbon emissions and energy costs for AI inference workloads in geo-distributed data centers},
  author={Hogade, Ninad and Pasricha, Sudeep},
  journal={IEEE Transactions on Sustainable Computing},
  volume={10},
  number={4},
  pages={628--641},
  year={2024},
  publisher={IEEE}
}

@article{nguyen2020generation,
  title={Generation expansion planning with renewable energy credit markets: A bilevel programming approach},
  author={Nguyen, Hieu T and Felder, Frank A},
  journal={Applied Energy},
  volume={276},
  pages={115472},
  year={2020},
  publisher={Elsevier}
}

@article{rosen1965existence,
  title={Existence and uniqueness of equilibrium points for concave n-person games},
  author={Rosen, J Ben},
  journal={Econometrica: Journal of the Econometric Society},
  pages={520--534},
  year={1965},
  publisher={JSTOR}
}
\end{document}